\documentclass[runningheads]{llncs}
\usepackage[T1]{fontenc}
\usepackage{graphicx}
\usepackage{listings}

\usepackage{longtable}

\begin{document}
\title{Experimental Toolkit for Manipulating Executable Packing}
%
%
\author{Alexandre D'Hondt\inst{1}\orcidID{0000-0001-6160-8624} \and
Charles Henry Bertrand Van Ouytsel\inst{2,3}\orcidID{0000-0001-5720-6569} \and
Axel Legay\inst{3}\orcidID{0000-0003-2287-8925}}
%
%
\institute{Universit\'e Catholique de Louvain, Rue Archimede 1, Louvain-la-Neuve, Belgium
\email{alexandre.dhondt@student.uclouvain.be,\{charles-henry.bertrand,axel.legay\}@uclouvain.be}\\
\url{http://www.uclouvain.be}}
\maketitle              
\begin{abstract}
Be it for a malicious or legitimate purpose, packing, a transformation that consists in applying various operations like compression or encryption to a binary file, i.e. for making reverse engineering harder or obfuscating code, is widely employed since decades already. Particularly in the field of malware analysis where a stumbling block is evasion, it has proven effective and still gives a hard time to scientists who want to efficiently detect it. While already extensively covered in the scientific literature, it remains an open issue especially when considering its detection time and accuracy trade-off. Many approaches, including machine learning, have been proposed but most studies often restrict their scope (i.e. malware and PE files), rely on uncertain datasets (i.e. built based on a super-detector or using labels from an questionable source) and do no provide any open implementation, which makes comparing state-of-the-art solutions tedious. Considering the many challenges that packing implies, there exists room for improvement in the way it is addressed, especially when dealing with static detection techniques. In order to tackle with these challenges, we propose an experimental toolkit, aptly called the Packing Box, leveraging automation and containerization in an open source platform that brings a unified solution to the research community and we showcase it with some experiments including unbiased ground truth generation, data visualization, machine learning pipeline automation and performance of open source packing static detectors.

\end{abstract}
\section{Introduction} 

\paragraph{Problem Statement}-- Packing detection has already been addressed in many researches since the early 2000's. Be it for the good or the bad, packing, the process of partially or completely transforming a binary file through operations such as compression or encryption, can be seen in many forms for many executable formats. Its use is not new, in particular with malicious software ; while antivirus products have flourished for identifying malware relying on signatures and heuristics, hackers have found ways to evade detection thanks to packing as it allows to hide the actual logic from matching it with signatures and patterns. However, packing is also used for legitimate purpose such as software protection and license management or simply for reducing the size.

The greatest dilemma in the challenge of detection is \textit{performance}, that is, the trade-off about maintaining enough accuracy while keeping the process achievable in a reasonable processing time. This could be addressed by finding a good balance between static and dynamic analysis. However, static analysis techniques can be as fast as a simple pattern matching while dynamic analysis results in a significant overhead because the target executable has to be executed. In order to detect packing, many static techniques were presented in the past, relying on entropy, signatures, control flow graphs or even ad hoc heuristics. Many high-quality researches have been published, often bringing excellent results to the forefront but never sharing a solution to repeat experiments and compare methods with each other. \textit{Repeatability} is thus another important concern. Moreover, machine learning has become the emerging trend for gaining in accuracy while keeping execution fast enough. Its optimal use, considering supervised learning, involves datasets of collected packed and not packed executables, but also capturing their most relevant characteristics, in a trade-off consisting of limiting the number of features and choosing the best ones for a reasonable performance while keeping the accuracy to a descent level. \textit{Feature engineering} is thus yet another concern. In order to reinforce this process, we can also leverage data visualization to better understand how packers work and hopefully find new relevant features. Furthermore, addressing machine learning requires making datasets that can be considered generalized enough, otherwise the resulting model will not cover every target executable. A major concern then lies in the \textit{relevance of} the available \textit{datasets} and the \textit{correctness} in the establishment of a ground truth considered for training models.

\paragraph{Research Questions}-- From the aforementioned considerations, we can see that there is room from improvement in packing detection. Therefore, the few following questions arise:

\begin{enumerate}
    \item How can we enhance \textit{repeatability} of the experiments of packing detection ?
    \item How can we provide a solution to build \textit{unbiased ground truths} ?
    \item What is the \textit{performance of state-of-the-art detectors} ?
    \item Can we leverage \textit{data visualization} to better understand what packers do ?
\end{enumerate}

\paragraph{Proposed solution}-- We want to contribute an experimental toolkit that automates dataset generation for building unbiased ground truths and provides tools for training machine learning models in an easy-to-configure and repeatable way. These tools will also include data visualization utilities and allow to assess the performance of open source detectors we can find in the wild.

The remainder of this paper first presents some background on the topic of packing detection, including common static techniques, ground truths generation methods and machine learning considerations, including related works. It then exposes our experimental toolkit, its capabilities and how it can help us solve our questions. Afterwards, we succinctly show some results of experiments we could lead with our solution. We finally conclude on what could be achieved and what remains to be further developed and studied.

\section{Background} \label{sec:background} 

\textit{Packing}, can be defined as set of transformations that modify the layout of a binary file (an executable, typically \texttt{.exe}, or an object, e.g. a DLL) without affecting its logic, adding a stub that allows to revert these transformations when executing the binary for restoring its logic in memory. In particular, this allows to hide binary's logic while residing on the disk, therefore preventing certain static detection systems from viewing the real logic (i.e. causing evasion of detection). But it can also be used to reduce its size or even to protect it against piracy.


\paragraph{Taxonomy}-- The related transformations are assumed to be independent but can be used in conjunction by packers. Among the possible transformations, we distinguish the followings:

\begin{itemize}
    \item \textit{Bundling} produces a single executable with multiple files. This is the typical behavior for portable versions of executables produced by tools like Cameyo, ThinApp or even Enigma Virtual Box.
    \item \textit{Compression} allows to reduce the size of the binary, in all or part, or even to obfuscate some regions of the binary. For instance, some very popular packers like UPX and UPack are especially designed for compressing binaries. Compression algorithms in use are typically LZMA, LZ77 or even Deflate.
    \item \textit{Encoding} is a reversible operation that changes a given character set to another, possibly of different size. Base64 and a simple XOR with a 1-byte key of value 1 are examples of encodings that could be used.
    \item \textit{Encryption} allows to obfuscate parts of the binary but also to protect the binary from piracy. It is a reversible operation that requires a key, either embedded or requested to the user when running the binary. For instance, Yoda Crypter encrypts sections on the spot, retaining the original layout.
    \item \textit{Mutation} alters the executable's code so that it uses a modified instruction set and architecture, e.g. using metamorphism (that is, the executable mutates its code at each execution).
    \item \textit{Protection} aims to make reversing the target binary harder, i.e. using anti-debugging, anti-tampering or other tricks that prevent debuggers or analysis tools to properly achieve reverse engineering tasks. For instance, Themida supports multiple anti-debugging protections including anti-virtualization.
    \item \textit{Virtualization} involves embedding a virtual machine that allows to virtualize executable's instructions for the sake of isolating it from the underlying operating system. This can be used for portability (e.g. BoxedApp Packer and Enigma Virtual Box) but for obfuscation as well (e.g. Molebox).
\end{itemize}

\paragraph{Static Techniques}-- Among the techniques for packing detection, static ones consist of actions performed without ever executing the binary (on the contrary of dynamic techniques which require execution). Some of them rely on characteristics of the executable formats and others on computed values. In the current literature, we find various methods relying on entropy, signatures, control flow graphs or heuristics (e.g. using a risk score). We note the followings:

\begin{itemize}
    \item \textit{Entropy}: Bintropy, presented in 2007 by Lyda \textit{et al.} \cite{lydaUsingEntropyAnalysis2007}, used block entropy measures (with a best block size of 256 bytes), either on the whole binary or per section, and applies thresholds for the average and highest block entropies. REMINDer, a tool proposed in 2009 by Han \textit{et al.} \cite{hanPackedPEFile2009}, used the distribution of entropy of the section containing the entry point as a criteria and outperformed some tools of that time like PEiD \cite{snakerPEiD2008}. Ugarte-Predrero \textit{et al.} \cite{ugartepedreroCounteringEntropyMeasure2012} introduced in 2012 a new metric called Entropy Surface Over a Threshold (ESOT), once again determining an optimal threshold to discriminate between packed and not packed binaries. In 2012 again, Naval \textit{et al.} \cite{navalESCAPEEntropyScore2012} proposed a fast method called ESCAPE using block entropies, focusing on blocks starting from the beginning of the disassembled code with incremental sizes up to a short value for limiting the processing time. All these methods achieved more than 95\% accuracy with their respective datasets.
    \item \textit{Signatures}: In 2011, Shin \textit{et al.} \cite{shinNewSignatureGeneration2011} proposed a signature format based on opcodes delimiting the unpacking algorithm (typical for a precise version of a packer), keeping the longest common sequence of opcodes in between for a given set of samples. In 2012, Naval \textit{et al.} \cite{navalSPADESignatureBased2012} presented a technique for generating signatures based on local alignment of byte sequences and a similarity score, empirically determining the threshold. In 2017, Saleh \textit{et al.} \cite{salehControlFlowGraphbased2017} published a new scheme relying on Control Flow Graphs (CFG) for generating signatures resilient to evasion, that is, able to detect different versions of the same packer with the same signature, even after the modification of some instructions. In 2017 as well, Hai \textit{et al.} \cite{haiPackerIdentificationBased2017} proposed another scheme based on metadata also relying on CFGs, using a frequency vector of occurrences of classified obfuscation techniques and identifying a packer via a chi-square test on the resulting metadata signature. All these approaches achieved a decent accuracy above 90\%.
    \item \textit{Control Flow Graphs (CFG)}: In 2010, Cesare \textit{et al.} \cite{cesareFastFlowgraphBased2010} proposed a classification system based on graph isomorphism and similarity for classifying polymorphic malware, either packed or not. They further improved their system in 2013 \cite{cesareMalwiseEffectiveEfficient2013}, combining two CFG-based algorithms (one matching exact strings while the other matched decompilation-based signatures). While not specifically oriented on packing, the method could apply for detecting packer variants. As presented here before for \textit{Signatures}, in 2017, Saleh \textit{et al.} \cite{salehControlFlowGraphbased2017} and Hai et al. \cite{haiPackerIdentificationBased2017} relied on CFGs for their signature schemes. In 2019, Li \textit{et al.} \cite{liConsistentlyExecutingGraphBasedApproach2019} proposed a novel approach relying on an alternative to CFGs, Consistently Executing Graphs (CEG), which could better model execution intents of the target binary, therefore better representing executables and using a custom distance for comparison with a reference collection of graphs. Most of these approaches achieved good results for detecting malware, yet considering packing as an open issue.
    \item \textit{Heuristics}: In 2008, Choi \textit{et al.} \cite{choiPEFileHeader2008} proposed PHAD (PE Header Analysis-based packing Detection) combining 8 structural features into a vector using an Euclidian distance with a custom threshold. In 2009, Treadwell \textit{et al.} \cite{treadwellHeuristicApproachDetection2009} proposed a heuristic detection approach relying on 8 static features of PE files pertaining to their loading into memory and used a risk score based on a weighted mean of per-feature risk scores, empirically setting weights. In 2013, Arora \textit{et al.} \cite{pareekHeuristicsbasedStaticAnalysis2013} used an approach similar to Treadwell \textit{et al.} this time focusing on 14 header-based features. In 2014, Saleh \textit{et al.} \cite{salehInstructionsBasedDetectionSophisticated2014} published a novel technique combining instructions-based features and 7 structural features. In 2015, Lim \textit{et al.} \cite{limMalEVEStaticDetection2015} proposed a model called Mal-EVE relying on 6 heuristic features combined into a risk score based on the power distance formula. All these approaches showed a decent accuracy, with Mal-EVE performing with 98,16\% but in 3,22 seconds while this of Arora \textit{et al.} performed with 99\% while keeping the processing time beneath 2 milliseconds.
\end{itemize}

Most of the aforementioned studies focused on malware analysis and Windows Portable Executables, often considering a single static technique and some others combining several of them. Almost all of them achieved good results on specific datasets and showed few or no comparison of performance with other tools or approaches, most often considering popular open tools like PEiD \cite{snakerPEiD2008}.

\paragraph{Ground Truths}-- Various sources of executables can be found in the wild, especially related to malware, but few provides packer labels or have labels determined based on proprietary or unpublished detection algorithms. It appears from many studies that some open datasets do not survive over time, making performance testing even more tedious with approaches from studies using more recent datasets. Among the interesting works for offering datasets, we note EMBER \cite{andersonEMBEROpenDataset2018} (2018, large collection of features from PE files), PackingData \cite{chesvectainPackingData2019} (2019, small collection of native PE files from Windows packed with 19 packers) and its extension \cite{dhondtDatasetPackedPE2021} (2021, 6 more packers added) but also a new similar set with ELF samples \cite{dhondtDatasetPackedELF2022} (2022, native Linux files packed with 6 different packers). We can also mention some former datasets that were used in some studies and that are now unmaintained like Malfease \cite{Malfease2008} (2008), VX Heaven \cite{VXHeaven2017} (2010 to 2017) or even ViruSign \cite{ViruSign2020} (stopped in 2020). Some more recent datasets are still actively maintained and usable, like Malware Bazaar \cite{MalwareBazaar2022} and VirusShare \cite{VirusShare2022}.

\paragraph{Machine Learning}-- This tends to become the best way to tie multiple techniques together, using their results as features to improve the models. While simpler techniques like custom heuristics use feature vectors with simple distance metrics to make predictions, machine learning algorithms offer far more possibilities in shaping models and leveraging features. We note the following related works for the different classes of learning algorithms:

\begin{itemize}
    \item \textit{Supervised learning}: In 2008, Perdisci \textit{et al.} \cite{perdisciMcBoostBoostingScalability2008} used a combination of 4 entropy-based, 5 structural and also $n$-grams ($n=2$ or $3$) features for training bagged decision trees (DT) and Multi-Layers Perceptrons (MLP) using a ground truth established using PEiD \cite{snakerPEiD2008}. In 2009, Shafiq \textit{et al.} \cite{shafiqPEProbeLeveragingPacker2009} used 189 PE structural features computable in real-time, focusing on malware detection and using 5 basic learning algorithms. In 2011, Wang \textit{et al.} proposed a generic packing detection framework relying on a feature vector of PE file header attributes, training a Support Vector Machine algorithm and selecting the best model using cross-validation. In 2014, Ugarte-Pedrero \textit{et al.} published a study based on a distance-based anomaly detection relying on 199 PE structural features and some classical learning algorithms including Random Forest (RF), SVM and MLP. In 2019, Biondi \textit{et al.} \cite{biondiEffectiveEfficientRobust2019} proposed a study based on 119 features (including composite ones), tree-based algorithms and relying on a ground truth acquired from a proprietary dataset from Cisco labeled with a superdetector (grouping 4 detectors), showing that a decrease in effectiveness could allow a significant increase in performance. In 2020, Aghakhani \textit{et al.} \cite{aghakhaniWhenMalwarePackin2020} proposed models relying on 543 features (PE headers and sections, strings and $n$-grams) using SVM, MLP and RF and the dataset EMBER \cite{andersonEMBEROpenDataset2018}. In 2021, Betrand Van Oytsel \textit{et al.} \cite{ouytselAnalysisMachineLearning2021} published a study identifying the most significant features and the best models among 11 learning algorithms. All the aforementioned approaches performed with an accuracy of more than 95\% on their respective datasets.
    \item \textit{Semi-supervised learning}: In 2011, Ugarte-Pedrero \textit{et al.} \cite{ugarte-pedreroSemisupervisedLearningPacked2011} proposed a new method for reducing the requirement of labeled data relying on header and section-based features but also entropy values, showing that 10\% of labeled data was sufficient to get good results. In 2012, Santos \textit{et al.} \cite{santosSemisupervisedLearningUnknown2011} iterated this approach by using collective classification, determining the optimal number of label instances required.
    \item \textit{Unsupervised learning}: In 2021, Noureddine \textit{et al.} \cite{noureddineSEPACSelfevolvingPacker2021} proposed a novel packer classifier leveraging cluster-based unsupervised learning with a particular distance metric combining different features for creating a system that could adapt to packer's evolutive nature.
\end{itemize}

From the aforementioned works, it appears that supervised learning was already extensively addressed while not identifying best features yet, except in the study of Betrand Van Oytsel \textit{et al.} \cite{ouytselAnalysisMachineLearning2021}. It is also worth being noted that no implementation for automating the learning pipeline and comparing models' results was provided yet.

\section{Toolkit} 

\begin{figure}[!htb]
    \centering
    \includegraphics[width=\linewidth]{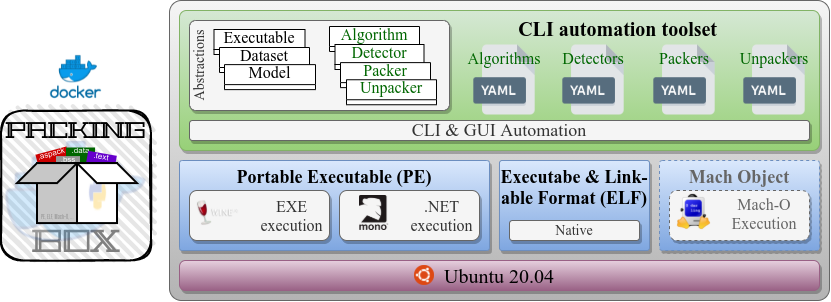}
    \caption{Architecture of the Docker image of the Packing Box}
    \label{fig:packing-box-architecture}
\end{figure}

For tackling with the drawbacks of the aforementioned approaches, we designed a container isolating the required logic to deal with datasets, features and learning algorithms. The \textbf{Packing Box} \cite{dhondtPackingBox2022}, presented in Figure \ref{fig:packing-box-architecture}, is architectured in several layers in a Docker image based on Ubuntu (lower layer) with libraries for supporting multiple executable formats (middle layer ; including PE \footnote{Portable Executable (Windows)}, ELF\footnote{Executable and Linkable Format (Linux)} and Mach-O\footnote{Mach Objects (MacOS)}) and our own framework containing our library and tools for automating packing-related resources (upper layer). Its data structures and abstractions allow for flexibility in integrating new algorithms, detectors, packers and unpackers by hand thanks to the convenient markup language, YAML\footnote{Yet Another Markup Language}.

While running the Docker image, the user ends up in a terminal that sets all the tooling available for tuning the box, including the \texttt{packing-box} (adminstrative utility for setting up and testing new items), \texttt{detector} (relies on the \textit{Detector} abstraction and allows to use multiple detectors at once as a superdetector or even to apply mass detection and to output statistics for a given dataset), \texttt{packer} and \texttt{unpacker} tools (rely on the \textit{Packer} and \textit{Unpacker} abstractions and allow to test the packing and unpacking processes or also allows to apply bulk packing/unpacking).

\paragraph{Items integration}-- The integrated software resources are already very varied, including 3 analyzers, 9 detectors, 16 packers and 3 unpackers (only counting the functional items ; the framework is in constant evolution and will integrate more items very soon). It also includes various utilities like Pefeats \cite{Pefeats2020} for feature extraction. Thanks to the abstractions (\textit{Packer}, \textit{Detector}, and so forth), definitions from the YAML files are translated into objects taking their user-defined parameters for tuning the execution. This allows to very conveniently define how to install packers, how to execute them or even set variants corresponding to multiple sets if inputs (e.g. tuning the compression level when using a compressing packer) or different packers' versions. The following abstractions are defined:

\begin{itemize}
    \item \textit{Analyzer}: It represents a file analyzer that outputs information but does not decide on packing. It is useful for inspecting binary file's characteristics. For instance, the old tool GetTyp \cite{phaxGetTyp2000} can inspect MS-DOS files or PortEx \cite{katjahahnPortEx2021} can inspect PE files.
    \item \textit{Detector}: This item represents a detector, that is, a tool that collects information like an \textit{Analyzer} but also outputs detection or suspicion, eventually deciding if the target binary is packed. For instance, Detect It Easy! \cite{horsicq2021}, a Python CLI version of PEiD \cite{dhondtPEiDCLI2021} or even RetDec \cite{avastRetDec2020} are integrated.
    \item \textit{Packer} and \textit{Unpacker}: These items represent packers and unpackers, callable from the console or through dataset generation with specific configurations. For instance, the very popular UPX is integrated (applies to any format), but also (for PE files) PEtite, Yoda Crypter, Kkrunchy, MEW (and many others) and (for Linux) Ezuri, GZEXE, MidgetPack (and some others).
\end{itemize}

Many parameters can be set such as the \textit{categories} (applies to packers and unpackers ; e.g. \textit{compressor}, \textit{protector}), the applicable \textit{formats} (e.g. PE32, ELF64), the \textit{install} procedure or the \textit{steps} for execution. For packers, the \textit{variants} parameter allows for easy declaration of either alternative configurations of inputs or alternative versions. Many examples of these definitions are available on the repository of the Packing Box \cite{dhondtPackingBox2022} and also explained in its documentation.

\begin{figure}[!htb]
    \centering
    \includegraphics[width=\linewidth]{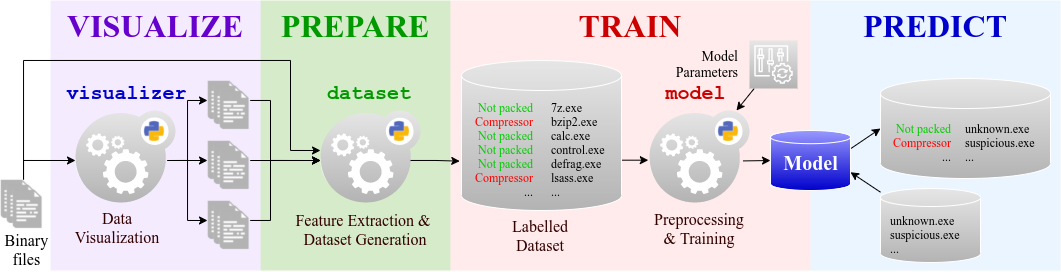}
    \caption{Machine Learning pipeline and the automation toolset}
    \label{fig:machine-learning-pipeline}
\end{figure}

Other tools are readily available to start with experiments, including \texttt{dataset}, \texttt{visualizer} and \texttt{model} that automate the steps of the machine learning pipeline as depicted on Figure \ref{fig:machine-learning-pipeline}. At the \textit{VISUALIZE} step, \texttt{visualizer} helps sort the input samples or even complete packer labels or dicard outliers. At the \textit{PREPARE} step, \texttt{dataset} provides a list of useful commands for creating, merging or splitting all kinds of datasets and simplify preparation for model training. Then at the \textit{TRAIN} step, \texttt{model} allows to train models and select the best parameters with a grid search based on the selected algorithm(s) as defined in the related declarative YAML file.

With these tools in hand, controlling the whole pipeline becomes a piece of cake, therefore drastically enhancing \textit{repeatability}.

\paragraph{Dataset generation}-- The \texttt{dataset} tool allows to manipulate datasets in many ways, including their creation, expansion, merging, splitting or even conversion to fileless datasets (only containing the features).

This tool relies on four abstractions:

\begin{itemize}
    \item \textit{Executable}: This provides attributes, including their metadata (filename, SHA256 hash, file type, ...), and properties, including features to be computed according to a selection of the available ones (hard coded for extraction but can also be derived based on the declarations from the related YAML file), for dealing with binary samples.
    \item \textit{Feature}: This represents a feature and its metadata (short name, description, source) and provides a binding with its YAML definition. This abstraction only applies to feature combination and derivation ; basic features depending on executable formats for extraction are hard coded.
    \item \textit{Dataset}: This enforces a given structure including dataset's metadata, the list of considered features (with their descriptions) and the files with their metadata. Thanks to this abstraction, many operations are defined in order to manipulate and combine datasets.
    \item \textit{FilelessDataset}: This allows to embed computed features instead of keeping the files which may occupy a large disk space. This way, datasets can be easily published and exchanged for a given set of features.
\end{itemize}

Thanks to this tool and the packer abstraction, we can build unbiased ground truths by providing a set of cleanware executables that will be packed by the Packing Box, therefore ensuring the right labels.

\paragraph{Data visualization}-- \texttt{dataset} also provides a way to inspect a single feature or a group of features together with their variance. It can plot features for visualizing the distribution of their values across labels (either packed/not-packed or per packer name). For the problem of outliers, a few utilities are also integrated for identifying them, namely through data visualization with the \texttt{visualizer} tool. This is especially aimed to plot binary files according to their sections with their entropy graphs. With this mechanics, we have a mean to act on the first steps of the learning pipeline, including feature engineering and dataset generation.

\paragraph{Model training}-- With the \texttt{model} tool, we complete the chain by providing a way to train models, after having tuned their hyper-parameters via their algorithms' YAML definitions (currently including 16 algorithms), either statically defined or for cross-validation. Using its convenient commands, \texttt{model} allows to train, test and inspect (including features ranking per importance) models and to collect performance metrics, for the sake of evaluating their efficiency and also comparing it with other models.

This tool relies on two abstractions:

\begin{itemize}
    \item \textit{Model}: Just like \textit{Dataset}, this enforces a structure for storing a model including its metadata, the list of considered features (with their descriptions) and its dump in Joblib format. It also contains a file holding metrics collected while testing or comparing with other models.
    \item \textit{DumpedModel}: This aims to encompass an orphan Joblib dump so that it can be used in the Packing Box, making possible to import dumped models from other studies.
\end{itemize}

This automation solves the issue of \textit{repeatability}, bringing clarity in the way parameters are defined and the conditions that lead to the best models.

\section{Experiments} 

For this paper, we focus our experiments on data visualization, performance evaluation of the integrated detectors and training machine learning models. The part related to feature engineering is still a work in progress. Before going further, we can run the Packing Box from the terminal with the following command:

\begin{lstlisting}[basicstyle=\footnotesize]
  $ docker run -it -h packing-box -v `pwd`:/mnt/share \
      packing-box
\end{lstlisting}

\paragraph{Entropy Visualization}-- Among the techniques based on entropy presented in Section \ref{sec:background}, we implemented Bintropy in a Python package \cite{dhondtBintropy2021} and added a visualization capability to plot binaries' structure and emphasize sections with their respective entropies. It is called from the \texttt{visualizer} tool integrated into the Packing Box. It provides a convenient way to visualize effects of some packing categories (i.e. compression, encryption and virtualization). Ultimately, it allows to deduce some features that may be useful with machine learning.

\begin{figure}[!htb]
    \centering
    \includegraphics[width=\linewidth]{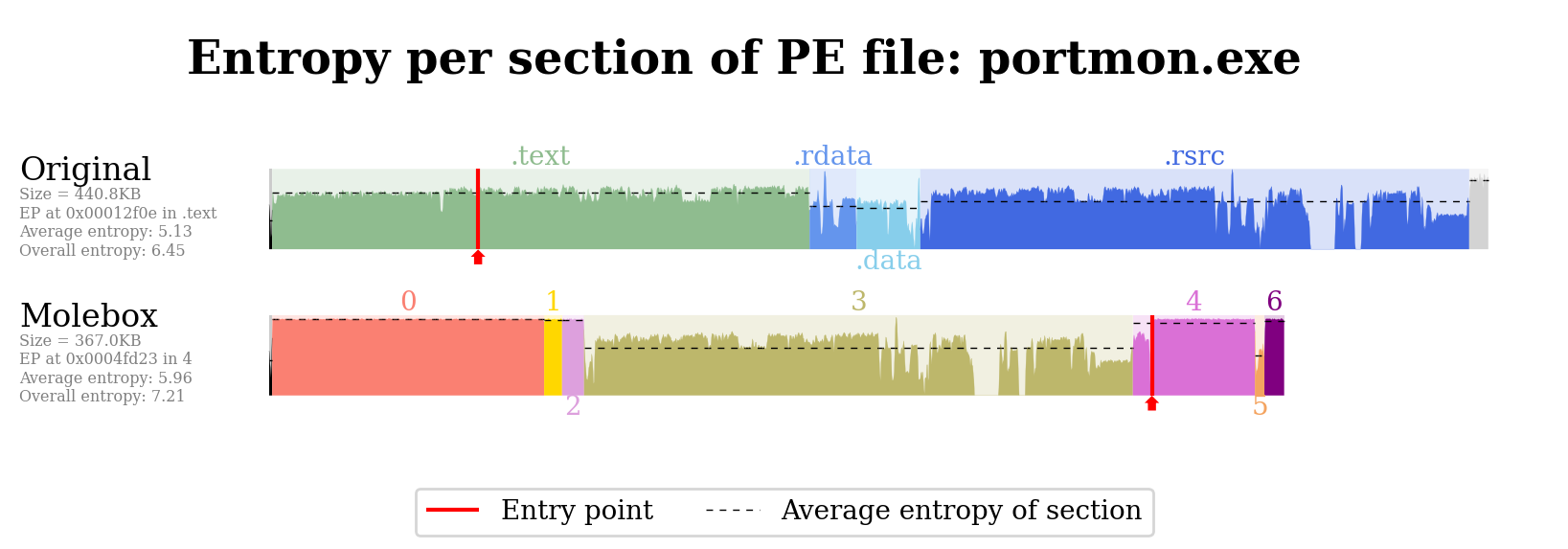}
    \caption{Example of visualization of a Windows PE packed with Enigma Virtual Box}
    \label{fig:packer-example-molebox}
\end{figure}

The command hereafter illustrates how to generate the plot from Figure \ref{fig:packer-example-molebox}.

\begin{lstlisting}[basicstyle=\footnotesize]
  packing-box$ visualizer plot 'portmon.exe' \
      dataset-packed-pe -l not-packed -l Molebox
\end{lstlisting}

This command consists of:

\begin{tabular}{@{}p{3.5cm}p{8cm}@{}}
    \texttt{plot} & command for plotting the entropy chart \\
    \texttt{portmon.exe} & target executable from the input dataset \\
    \texttt{dataset-packed-pe} & target folder of the input dataset (e.g. \cite{dhondtDatasetPackedPE2021}) \\
    \texttt{-l ...} & labels to be selected from the tree structure within the folder of the input dataset \\
\end{tabular}\\

The example in Figure \ref{fig:packer-example-molebox} shows a native Windows executable in its original form and its packed version processed with Molebox, a virtualizer. This plot allows to see multiple effects of this packer ; sections are renamed and compressed (possibly even encrypted) except the one containing the resources (\texttt{.rsrc}) and a few new sections containing the virtual machine are appended. Moreover, we see that it strips the overlay (the gray zone at the end of the original binary, which is a part unreferenced in the section table).

\begin{figure}[!htb]
    \centering
    \includegraphics[width=\linewidth]{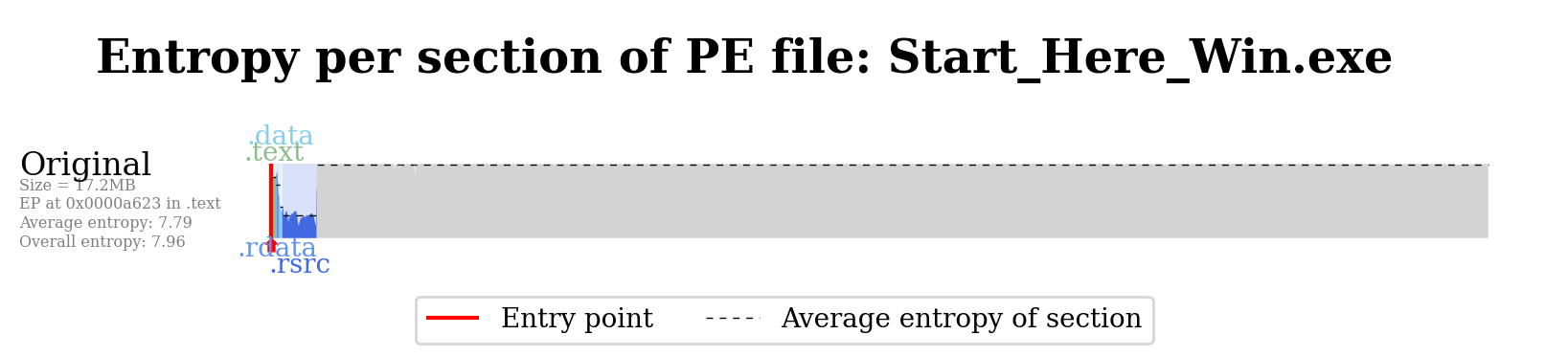}
    \caption{Example of visualization of a Windows PE containing data in the overlay}
    \label{fig:example-loaded-overlay}
\end{figure}

The example in Figure \ref{fig:example-loaded-overlay} shows a native Windows executable in its original form containing massive data in its overlay (the long section in gray that represents 16 megabytes out of 17 of the executable, typical for an installer). It shows its importance as many packers strip the overlay and therefore may prevent this kind of executable from working after packing.

In further experiments, if we want to label our datasets with the categories as described in Section \ref{sec:background}, we can first inspect the homepages of the related packers to determine from their descriptions which categories they relate to. We can further refine these labels by visually inspecting samples with the help of the \texttt{visualizer} tool. Note that, in the example of Figure \ref{fig:packer-example-molebox}, the homepage of Molebox explicitly mentions the categories of compression, encryption and virtualization, which are confirmed by its visualization.

\paragraph{Detectors Performance}-- Unfortunately, given the fact that many studies do not provide an open implementation, it is difficult at this point to compare the performance of the various techniques presented in Section \ref{sec:background}. Note that we could implement many of them on our own but, for sure, it requires a large amount of development time. However, we could find a few detectors in the wild (especially on GitHub) including static heuristics relying on entropy or structural characteristics of executables formats (mostly PE). The characteristics of the in-scope detectors are presented in Table \ref{tab:detectors-characteristics} of Appendix \ref{app:detectors-characteristics}.

We thus led an experiment on these detectors based on our first dataset \cite{dhondtDatasetPackedPE2021} (which is presented in Table \ref{tab:reference-dataset} of Appendix \ref{app:reference-dataset}) to compare their accuracy, processing time and time complexity. Note that this experiment currently considers the PE format but also applies to ELF with our second dataset \cite{dhondtDatasetPackedELF2022}.

\begin{itemize}
    \item \textit{Accuracy}: We divided our experiments into 4 classes, taking 2 parameters into account: whether classification is binary (packed or not) or multi-class (per packer name) and if detection is strong (the detector outputs either a formal decision or detection traces that lead to a formal decision given a simplistic heuristic of our own) or weak (the output provides detection traces but also suspicions that may yield a decision that is less robust). With the results of the 4 classes, we then considered the lower and upper bounds. Accuracy metrics were collected based on the execution of each detector on all the subsets (not packed and 25 packers) from our dataset \cite{dhondtDatasetPackedPE2021}, leveraging the \texttt{detector} tool from the Packing Box for bulk detection.
    \item \textit{Processing time}: This was evaluated by running the detectors against the samples from our dataset \cite{dhondtDatasetPackedPE2021}, only keeping the results for file sizes from 0 to 10 megabytes. We then considered the lower and upper bounds. 
    \item \textit{Time complexity}: For this, we took the results from the evaluation of the processing time and applied curve fitting to get the best fit among constant, logarithmic, linear, linear-algorithmic and quadratic complexities.
\end{itemize}

The following command illustrates how to get the accuracy for a given dataset using the \texttt{detector} tool of the Packing Box. While scripted to bulk-process every detector on each subset of our reference dataset \cite{dhondtDatasetPackedPE2021}, we can obtain the accuracy for the 4 different classes.

\begin{lstlisting}[basicstyle=\footnotesize]
  packing-box$ detector dataset-packed-pe -d die -m [-b] [-w]
\end{lstlisting}

This command consists of:

\begin{tabular}{@{}p{3.5cm}p{8cm}@{}}
    \texttt{dataset-packed-pe} & target folder of the input dataset (e.g. \cite{dhondtDatasetPackedPE2021}) \\
    \texttt{-d ...} & integrated detector to be selected \\
    \texttt{-m} & enable metrics acquisition (i.e. accuracy) \\
    \texttt{-b} & enable binary classification \\
    \texttt{-w} & enable weak mode \\
\end{tabular}\\

The results of our experiment are presented in Table \ref{tab:detectors-performance}. This shows the different metrics collected according to the aforementioned methodology and the related observations we could make in terms of detection techniques and their performance.

\begingroup
\renewcommand*{\arraystretch}{1.45}
\begin{longtable}{@{}p{1.8cm}p{1.5cm}p{1.2cm}p{1.6cm}p{5cm}@{}}
\multicolumn{1}{l}{\bfseries Detector}  & \multicolumn{1}{l}{\bfseries Accuracy} & \multicolumn{1}{l}{\bfseries Time} & \multicolumn{1}{l}{\bfseries Complexity} & \multicolumn{1}{l}{\bfseries Observations} \\
Bintropy \cite{dhondtBintropy2021} & 58,71\%                        & 0,5--10,5s  & $\mathcal{O}(n)$      & Solely relies on entropy (i.e. block entropy). \\
DIE \cite{horsicq2021} & 84,25\%--85,81\%               & ~450ms     & $\mathcal{O}(log(n))$ & Stable detection, null FPR, strong signatures combined with some heuristics (but that do not necessarily compensate missing signatures). \\
Manalyze \cite{ivankwiatkowskiManalyze2014} & 45,17\%--94,83\%               & ~350ms     & $\mathcal{O}(log(n))$ & Unstable detection, high FPR, relying on a short list of known section names and some simplistic heuristics. \\
PEFrame \cite{guelfowebPEFrame2019} & 57,42\%--87,52\%              & 1,39--27,6s & $\mathcal{O}(n)$      & Flawed ; only relies on signatures and entropy. \\
PEiD \cite{dhondtPEiDCLI2021} & 69,29\%--71,63\%               & 1,69--2,37s & $\mathcal{O}(n)$      & Stable detection, low FPR, database misses many signatures. \\
PEPack \cite{PEPack2017} & out-of-scope & 300ms      & $\mathcal{O}(1)$      & Not tested, relies on the same signatures DB as the original PEiD (no added value). \\
PyPacker-Detect \cite{dhondtPyPackerDetectRefactored2021} & 79,69\%--94,65\%               & 0,71--1,36s & $\mathcal{O}(n)$      & Decent FPR, provides good results and its heuristics cover cases not covered by the PEiD signatures. \\
PyPEiD \cite{ffriPyPeid2020} & out-of-scope & 0,82--2,26s & $\mathcal{O}(n)$      & Not tested, but could be in a near future as it augments PEiD signatures with Yara rules. \\
RetDec \cite{avastRetDec2020} & 70,71\%--75,76\%               & 0,32--0,84s & $\mathcal{O}(n)$      & Relatively stable detection, low FPR, relying on custom signatures and some heuristics covering unmatched cases. \\
\caption{Performance test results for in-scope detectors}\label{tab:detectors-performance}
\end{longtable}
\endgroup

In this table, we note that a few detectors have a stable detection (including DIE, PEiD and RetDec) while others do not (Manalyze, PEFrame and PyPackerDetect). We also see that only a few ones perform under 1 second (the best performers being DIE, Manalyze and PEPack and the worst ones being PEFrame and Bintropy). Moreover, we point out that most of the detectors have a complexity in $\mathcal{O}(n)$ which means they parse the entire binary. While this result is logical for Bintropy (as it requires to compute the entropy on the whole binary), for some others it is not. PEiD, which was used matching signatures from the Entry Point, surprisingly holds this complexity while we expected $\mathcal{O}(1)$ because matching the signature from the Entry Point requires a constant calculation and a simple matching with a pattern that is bounded to a constant maximum length. However, this result comes from the underlying library that systematically parses the entire file when called. It is also worth being noted that even the best detector, DIE, only performs at around 86\%. This is due to the specificity of the dataset in use, which contains multiple recent custom packers found on GitHub like Alienyze (which may have failed packing because of its demo version), Amber and Eronana's Packer.

From these results, we see that DIE is the best performer in terms of accuracy, followed by PyPackerDetect. It shall be noted however that PyPackerDetect holds a non-negligible FPR. Regarding time complexities, DIE remains the best. In terms of processing time, also considering the FPR and accuracy range, once again, DIE is at the top. In terms of candidates for the superdetector, Bintropy and Manalyze could be excluded for their accuracy. Moreover, PeFrame has a particularly high cost in processing time and, unless it performs very well after re-evaluation, it could then be discarded. Among the best performers, we thus keep DIE, PEiD, PyPackerDetect and RetDec. PyPeid could be evaluated against PEiD to see if it would not be a better match.

\paragraph{Learning Models}-- This experiment is aimed to showcase the automation of the learning pipeline. Other experiments for identifying the best features and models and trying to identify packer categories are left for future works. Following the learning pipeline of Figure \ref{fig:machine-learning-pipeline}, we already showcased data visualization in our first experiment. We now address dataset generation and model training, that is, the PREPARE, TRAIN and TEST phases.

The following commands illustrates how to make a dataset from our reference one \cite{dhondtDatasetPackedPE2021}, selecting some not-packed and UPX-packed samples and then adding our own UPX-packed samples using the packer's integration in the Packing Box.

\begin{lstlisting}[basicstyle=\footnotesize]
  packing-box$ dataset update dataset-upx -s \
      dataset-packed-pe/not-packed
  packing-box$ dataset update dataset-upx -s \
      dataset-packed-pe/packed/UPX -l labels.json
  packing-box$ dataset make dataset-upx -n 300 -f PE \
      -p upx --pack-all
\end{lstlisting}

These commands consist of:

\begin{tabular}{@{}p{2.7cm}p{8.8cm}@{}}
    \texttt{update} & command for updating a dataset with input samples (creates the dataset if it does not exist) \\
    \texttt{make} & command for packing samples with the Packing Box \\
    \texttt{dataset-upx} & new dataset \\
    \texttt{-s ...} & source folder to get samples from \\
    \texttt{-l ...} & JSON file containing the labels to be used (format: key is a SHA256 hash, value is the target label) \\
    \texttt{-n} & number of samples to be generated \\
    \texttt{-f} & executable format to be selected from the samples of the source folder \\
    \texttt{-p} & packer to be selected \\
    \texttt{-{-}pack-all} & enable packing every sample (otherwise, samples are packed or not so that the dataset gets balanced) \\
\end{tabular}\\

This set of commands completes the PREPARE phase of the pipeline. We now have a dataset consisting of 854 samples whose 422 are UPX-packed (this can be inspected using \texttt{dataset show dataset-upx}). We can then start training models.

The following command trains a model using the decision tree algorithm based on the dataset we just made.

\begin{lstlisting}[basicstyle=\scriptsize]
  packing-box$ model train dataset-upx -a dt
  <<<snipped>>>
  Name: dataset-upx\_pe\_854\_dt\_f132
  -----  --------  ---------  -------  ---------  -------  -------
  .      Accuracy  Precision  Recall   F-Measure  MCC      AUC
  Train  100.00%   100.00%    100.00%  100.00%    100.00%  100.00%
  Test   100.00%   100.00%    100.00%  100.00%    100.00%  100.00%
\end{lstlisting}

We obtain a perfectly fitted model that gets named according to the reference dataset and its related size, the learning algorithm used and the number of applicable features (that is, excluding the features that got a null variance). With this command executed, we completed the TRAIN phase.

We can now go to the TEST phase and apply our new model to another dataset. For this purpose, we can create another custom dataset, an unbiased ground truth, by using \texttt{dataset make} without the \texttt{-{-}pack-all} option, e.g. with 500 samples (by default, taken from within the Packin Box).

The following command tests our new model based on our new test dataset.

\begin{lstlisting}[basicstyle=\scriptsize]
  packing-box$ model test dataset-upx\_pe\_854\_dt\_f132 dataset-upx-test
  <<<snipped>>>
  Test results for: dataset-upx-test
  --------  ---------  -------  ---------  -------  -------
  Accuracy  Precision  Recall   F-Measure  MCC      AUC    
  100.00%   100.00%    100.00%  100.00%    100.00%  100.00%
\end{lstlisting}

The result shows that our model perfectly performs with the test dataset or deciding whether samples are packed or not. In a few commands, we thus followed the complete learning pipeline. By playing with input datasets, it is now easy to study other classifications, i.e. packer categories.

\section{Conclusion} 

\paragraph{Summary}-- In this paper, we first stated the background and enumerated techniques of static detection of executable packing as currently described in the literature, including the latest advances in this field related to machine learning. We then briefly presented the architecture of our experimental toolkit, the Packing Box, and the capabilities it offers. Finally, we led some experiments for showcasing the functionalities we included and the way the complete machine learning pipeline could be walked through in only a few terminal commands.

\paragraph{Contributions}-- We designed a platform for studying executable packing, not especially focused on malware analysis as usual and limited to Windows PE but also applicable to other executable formats like ELF or Mach-O, and bringing an open source unified solution to the research community that addresses the many related practical challenges everything at once. With the Packing Box, we have shown that it significantly enhances repeatability in our experiments as it fully integrates assets (including detectors and packers) and it automates the complete machine learning pipeline (from dataset generation to model training, testing and comparison). Thanks to the integration of packers, we could show that it allows to generate unbiased ground truths starting from a set of cleanware samples. Thanks to the integration of detectors, we could bulk-process a dataset we made in order to evaluate the performance of state-of-the-art detectors and we could even plan to use the best ones together as a super-detector. Moreover, thanks to the additional toolset included in the Packing Box, we could also apply data visualization to better understand some categories of packers. With this mechanics in hand, we now have a complete solution to compare and enhance machine learning models for the static detection of executable packing which makes this experimental toolkit very promising.

\paragraph{Future Works}-- We already planned a few enhancements in terms of functionalities but also in our experiments as well ;

\begin{itemize}
    \item \textit{Support for executable formats}: While the PE and ELF formats are already fully supported and respectively largely and moderately covered by the integrated assets (i.e. detectors and packers), the Mach-O format still lacks integration (because of installation issues of the underlying library on the platform). This technical issue is currently being addressed.
    \item \textit{Packer categories}: We mentioned categories at the beginning of Section \ref{sec:background} but we did not present any experiment focused on detecting them as this is still a work in progress. We plan to train many models based on many datasets of our own in the hope of identifying sets of features that could better apply to one category or another.
    \item \textit{Feature engineering}: We already iterated the logic for handling features and have finally decided to also handle them in the declarative YAML format. Throughout our review of the literature, we gathered state-of-the-art features and established our own in what we called the \textit{Feature Book}. We thus plan to implement this to further refine our features set for our next experiments.
    \item \textit{Implementation of state-of-the-art techniques}: From the techniques mentioned in Section \ref{sec:background}, we want to implement a few of them in order to provide a comparison and to measure how well our approach could perform better.
\end{itemize}

\bibliographystyle{splncs04}
\bibliography{bibliography}

\begin{thebibliography}{10}
\providecommand{\url}[1]{\texttt{#1}}
\providecommand{\urlprefix}{URL }
\providecommand{\doi}[1]{https://doi.org/#1}

\bibitem{aghakhaniWhenMalwarePackin2020}
Aghakhani, H., Gritti, F., Mecca, F., Lindorfer, M., Ortolani, S., Balzarotti,
  D., Vigna, G., Kruegel, C.: When malware is packin' heat; limits of machine
  learning classifiers based on static analysis features. In: NDSS. p.~20
  (2020). \doi{10.14722/ndss.2020.24310}

\bibitem{andersonEMBEROpenDataset2018}
Anderson, H.S., Roth, P.: {{EMBER}}: {{An}} open dataset for training static
  {{PE}} malware machine learning models. ArXiv e-prints
  \textbf{abs/1804.04637} (2018), \url{https://arxiv.org/abs/1804.04637}

\bibitem{avastRetDec2020}
{Avast}: {{RetDec}}, \url{https://github.com/avast/retdec}

\bibitem{biondiEffectiveEfficientRobust2019}
Biondi, F., Enescu, M.A., Given-Wilson, T., Legay, A., Noureddine, L., Verma,
  V.: Effective, efficient, and robust packing detection and classification.
  Computers \& Security  \textbf{85},  436--451 (2019).
  \doi{10.1016/j.cose.2019.05.007}

\bibitem{cesareFastFlowgraphBased2010}
Cesare, S., Xiang, Y.: A {{Fast Flowgraph Based Classification System}} for
  {{Packed}} and {{Polymorphic Malware}} on the {{Endhost}}. In: AINA. pp.
  721--728. {IEEE} (2010). \doi{10.1109/AINA.2010.121}

\bibitem{cesareMalwiseEffectiveEfficient2013}
Cesare, S., Xiang, Y., Zhou, W.: Malwise - {{An Effective}} and {{Efficient
  Classification System}} for {{Packed}} and {{Polymorphic Malware}}. IEEE
  Transactions on Computers  \textbf{62}(6),  1193--1206 (2013).
  \doi{10.1109/TC.2012.65}

\bibitem{chesvectainPackingData2019}
{chesvectain}: {{PackingData}},
  \url{https://github.com/chesvectain/PackingData}

\bibitem{choiPEFileHeader2008}
Choi, Y.S., Kim, I.K., Oh, J.T., Ryou, J.C.: {{PE File Header Analysis-Based
  Packed PE File Detection Technique}} ({{PHAD}}). In: ISCSA. pp. 28--31.
  {IEEE} (2008). \doi{10.1109/CSA.2008.28}

\bibitem{ffriPyPeid2020}
{FFRI}: {{PyPeid}}, \url{https://github.com/FFRI/pypeid}

\bibitem{guelfowebPEFrame2019}
{guelfoweb}: {{PEFrame}}, \url{https://github.com/guelfoweb/peframe}

\bibitem{haiPackerIdentificationBased2017}
Hai, N.M., Ogawa, M., Tho, Q.T.: Packer {{Identification Based}} on {{Metadata
  Signature}}. In: SSPREW. p.~11. {ACM} (2017). \doi{10.1145/3151137.3160687}

\bibitem{hanPackedPEFile2009}
Han, S., Lee, K., Lee, S.: Packed {{PE File Detection}} for {{Malware
  Forensics}}. In: ICCSA. pp.~1--7. {IEEE} (2009). \doi{10/bm6p4g}

\bibitem{horsicq2021}
{horsicq}: {{DIE}}, \url{https://github.com/horsicq/DIE-engine/releases}

\bibitem{ivankwiatkowskiManalyze2014}
{Ivan Kwiatkowski}: Manalyze, \url{https://github.com/JusticeRage/Manalyze}

\bibitem{katjahahnPortEx2021}
{katjahahn}: {{PortEx}}, \url{https://github.com/katjahahn/PortEx}

\bibitem{liConsistentlyExecutingGraphBasedApproach2019}
Li, X., Shan, Z., Liu, F., Chen, Y., Hou, Y.: A {{Consistently-Executing
  Graph-Based Approach}} for {{Malware Packer Identification}}. IEEE Access
  \textbf{7},  51620--51629 (2019). \doi{10.1109/ACCESS.2019.2910268}

\bibitem{limMalEVEStaticDetection2015}
Lim, C., {Nicsen}: Mal-{{EVE}}: {{Static}} detection model for evasive malware.
  In: CHINACOM. pp. 283--288. {IEEE} (2015). \doi{10/ghm894}

\bibitem{lydaUsingEntropyAnalysis2007}
Lyda, R., Hamrock, J.: Using {{Entropy Analysis}} to {{Find Encrypted}} and
  {{Packed Malware}}. IEEE Security Privacy  \textbf{5}(2),  40--45 (2007).
  \doi{10.1109/MSP.2007.48}

\bibitem{Malfease2008}
Malfease,
  \url{https://web.archive.org/web/20141221153307/http://malfease.oarci.net}

\bibitem{MalwareBazaar2022}
{{MalwareBazaar}}, \url{https://bazaar.abuse.ch/browse}

\bibitem{navalESCAPEEntropyScore2012}
Naval, S., Laxmi, V., Gaur, M.S., Vinod, P.: {{ESCAPE}}: {{Entropy Score
  Analysis}} of {{Packed Executable}}. In: SIN. pp. 197--200. {ACM} (2012).
  \doi{10.1145/2388576.2388607}

\bibitem{navalSPADESignatureBased2012}
Naval, S., Laxmi, V., Gaur, M.S., Vinod, P.: {{SPADE}}: {{Signature Based
  PAcker DEtection}}. In: SecurIT. pp. 96--101. {{SecurIT}} '12, {ACM} (2012).
  \doi{10.1145/2490428.2490442}

\bibitem{noureddineSEPACSelfevolvingPacker2021}
Noureddine, L., Heuser, A., Puodzius, C., Zendra, O.: {{SE-PAC}}: {{A}}
  self-evolving packer classifier against rapid packers evolution. In: CODASPY.
  pp. 281--292. {{CODASPY}} '21, {Association for Computing Machinery} (2021).
  \doi{10.1145/3422337.3447848}

\bibitem{ouytselAnalysisMachineLearning2021}
Ouytsel, C.H.B.V., Given-Wilson, T., Minet, J., Roussieau, J., Legay, A.:
  Analysis of machine learning approaches to packing detection. CoRR
  \textbf{abs/2105.00473} (2021), \url{https://arxiv.org/abs/2105.00473}

\bibitem{pareekHeuristicsbasedStaticAnalysis2013}
Pareek, H., Arora, R., Singh, A.: A {{Heuristics-based Static Analysis
  Approach}} for {{Detecting Packed PE Binaries}}. IJSIA  \textbf{7},  257--268
  (2013). \doi{10.14257/ijsia.2013.7.5.24}

\bibitem{Pefeats2020}
Pefeats, \url{https://bit.ly/3CCRv8a}

\bibitem{PEPack2017}
{{PEPack}}, \url{https://github.com/merces/pev}

\bibitem{perdisciMcBoostBoostingScalability2008}
Perdisci, R., Lanzi, A., Lee, W.: {{McBoost}}: {{Boosting}} scalability in
  malware collection and analysis using statistical classification of
  executables. In: ACSAC. pp. 301--310 (2008). \doi{10.1109/ACSAC.2008.22}

\bibitem{phaxGetTyp2000}
{PHaX}: {{GetTyp}}, \url{https://www.helger.com/gt/gt.htm}

\bibitem{salehControlFlowGraphbased2017}
Saleh, M., Ratazzi, E.P., Xu, S.: A control flow graph-based signature for
  packer identification. In: MILCOM. pp. 683--688. {IEEE} (2017).
  \doi{10.1109/MILCOM.2017.8170793}

\bibitem{salehInstructionsBasedDetectionSophisticated2014}
Saleh, M., Ratazzi, E.P., Xu, S.: Instructions-{{Based Detection}} of
  {{Sophisticated Obfuscation}} and {{Packing}}. In: MILCOM. pp.~1--6. {IEEE}
  (2014). \doi{10.1109/MILCOM.2014.9}

\bibitem{santosSemisupervisedLearningUnknown2011}
Santos, I., Nieves, J., Bringas, P.G.: Semi-supervised {{Learning}} for
  {{Unknown Malware Detection}}. In: Abraham, A., Corchado, J.M., González,
  S.R., De~Paz~Santana, J.F. (eds.) DCAI. pp. 415--422. {Springer} (2011),
  \url{https://link.springer.com/chapter/10.1007/978-3-642-19934-9\_53}

\bibitem{shafiqPEProbeLeveragingPacker2009}
Shafiq, M.Z., Tabish, S.M., Farooq, M.: {{PE-Probe}}: {{Leveraging Packer
  Detection}} and {{Structural Information}} to {{Detect Malicious Portable
  Executables}}. In: VB. vol.~8, p.~10 (2009), \url{https://bit.ly/3craRSM}

\bibitem{shinNewSignatureGeneration2011}
Shin, D., Im, C., Jeong, H., Kim, S., Won, D.: The new signature generation
  method based on an unpacking algorithm and procedure for a packer detection.
  In: IJAST. vol.~27, pp. 59--78. {IJAST} (2011),
  \url{https://www.earticle.net/Article/A147420}

\bibitem{snakerPEiD2008}
{snaker}: {{PEiD}}, \url{https://github.com/wolfram77web/app-peid}

\bibitem{treadwellHeuristicApproachDetection2009}
Treadwell, S., Zhou, M.: A heuristic approach for detection of obfuscated
  malware. In: ISI. pp. 291--299. {IEEE} (2009). \doi{10.1109/ISI.2009.5137328}

\bibitem{ugarte-pedreroSemisupervisedLearningPacked2011}
Ugarte-Pedrero, X., Santos, I., Bringas, P.G., Gastesi, M., Esparza, J.M.:
  Semi-supervised learning for packed executable detection. In: ICNSS. pp.
  342--346. {IEEE} (2011). \doi{10.1109/ICNSS.2011.6060027}

\bibitem{ugartepedreroCounteringEntropyMeasure2012}
Ugarte-Pedrero, X., Santos, I., Sanz, B., Laorden, C., Bringas, P.G.:
  Countering entropy measure attacks on packed software detection. In: CCNC.
  pp. 164--168. {IEEE} (2012). \doi{10.1109/CCNC.2012.6181079}

\bibitem{ViruSign2020}
{{ViruSign}}, \url{https://www.virusign.com}

\bibitem{VirusShare2022}
{{VirusShare}}, \url{https://virusshare.com}

\bibitem{VXHeaven2017}
{{VX Heaven}},
  \url{https://web.archive.org/web/20170817143838/http://vxheaven.org/}

\bibitem{dhondtBintropy2021}
XXX: Bintropy, \url{https://anonymous.4open.science/r/bintropy-E18F}

\bibitem{dhondtDatasetPackedELF2022}
XXX: Dataset of packed elf files,
  \url{https://anonymous.4open.science/r/dataset-packed-elf-0602}

\bibitem{dhondtDatasetPackedPE2021}
XXX: Dataset of packed pe files,
  \url{https://anonymous.4open.science/r/dataset-packed-pe-06E8}

\bibitem{dhondtPackingBox2022}
XXX: Packing-{{Box}},
  \url{https://anonymous.4open.science/r/docker-packing-box-055A}

\bibitem{dhondtPEiDCLI2021}
XXX: {{PEiD}} ({{CLI}}), \url{https://anonymous.4open.science/r/peid-ED0F}

\bibitem{dhondtPyPackerDetectRefactored2021}
XXX: {{PyPackerDetect}} (fork),
  \url{https://anonymous.4open.science/r/PyPackerDetect-CC0E}

\end{thebibliography}

\appendix
\addcontentsline{toc}{chapter}{APPENDICES}

\section{Characteristics of Detectors}\label{app:detectors-characteristics}

\begingroup
\renewcommand*{\arraystretch}{1.45}
\begin{longtable}{@{}p{.14\linewidth}p{.13\linewidth}p{.04\linewidth}p{.04\linewidth}p{.04\linewidth}p{.59\linewidth}@{}}
\multicolumn{1}{l}{\bfseries Detector}  & \multicolumn{1}{l}{\bfseries Formats} & \rotatebox{60}{\bfseries Multiclass} & \rotatebox{60}{\bfseries Weak Mode} & \rotatebox{60}{\bfseries Superdetector} & \multicolumn{1}{l}{\bfseries Comments} \\
Bintropy \cite{dhondtBintropy2021} & PE, ELF,\newline MSDOS & N & N & N & Python CLI version of the tool of Lyda \textit{et al.} \cite{lydaUsingEntropyAnalysis2007} that integrates a plot capability. \\
DIE \cite{horsicq2021} & All & Y & N & Y & Most diversified detector we could find, including support for PE, ELF and Mach-O ; it contains many fine-grained custom signatures and heuristics and is used in many studies. \\
Manalyze \cite{ivankwiatkowskiManalyze2014} & PE,\newline MSDOS     & Y & Y & N & Open source tool that relies on PEiD's signatures DB, Yara rules and some heuristics (e.g. pattern matching for a small set of supported packers), sometimes used in studies. \\
PEFrame \cite{guelfowebPEFrame2019} & PE,\newline MSDOS     & Y & Y & N & Open source tool that relies on PEiD's signatures DB and some heuristics, found in only one study. \\
PEiD \cite{dhondtPEiDCLI2021} & PE,\newline MSDOS & Y & N & Y & Python CLI version of the popular tool that also integrates a basic signature generation capability and holds a sanitized database (with an aggregation of many sources found on GitHub repositories and cleaned up from its signatures unrelated to packing) that we implemented for use with the Packing Box. \\
PEPack \cite{PEPack2017} & PE,\newline MSDOS     & Y & N & N & Open source tool part of the unmaintained \texttt{pev} library and relying on PEiD's signatures DB. \\
PyPacker-Detect \cite{dhondtPyPackerDetectRefactored2021} & PE,\newline MSDOS     & Y & Y & Y & Fork of the unmaintained open source tool that relies on PEiD's signatures DB and a few simple heuristics. \\
PyPEiD \cite{ffriPyPeid2020} & PE,\newline MSDOS     & Y & N & N & Other Python implementation of PEiD, relying on its signatures DB also integrating Yara rules. \\
RetDec \cite{avastRetDec2020} & All          & Y & N & Y & Open source retargetable machine-code decompiler supporting many executable file format and using custom signatures and header-based common heuristics for packing detection. \\
\caption{Base characteristics of in-scope detectors}\label{tab:detectors-characteristics}
\end{longtable}
\endgroup

\section{Reference Dataset}\label{app:reference-dataset}

\begingroup
\renewcommand*{\arraystretch}{1.3}
\begin{longtable}{@{}p{.35\textwidth}p{.25\textwidth}p{.15\textwidth}p{.25\textwidth}@{}}
\multicolumn{1}{l}{\textbf{Name}} & \multicolumn{1}{l}{\textbf{\#Executables}} & \multicolumn{1}{l}{\textbf{Size}} & \multicolumn{1}{l}{\textbf{Formats}} \\
Alienyze & 126 & 67MB & .NET,PE32 \\
Amber & 139 & 81MB & PE32,PE64 \\
ASPack & 118 & 39MB & PE32 \\
BeroEXEPacker & 116 & 31MB & MSDOS \\
Enigma VirtualBox & 123 & 133MB & PE32 \\
Eronana's Packer & 139 & 37MB & .NET,PE32 \\
Exe32pack & 129 & 3MB & MSDOS \\
EXpressor & 121 & 23MB & PE32 \\
FSG & 117 & 23MB & .NET,PE32 \\
JDPack & 120 & 30MB & .NET,PE32 \\
MEW & 116 & 32MB & MSDOS \\
Molebox & 119 & 26MB & PE32 \\
MPRESS & 116 & 23MB & MSDOS,PE32 \\
Neolite & 116 & 61MB & PE32 \\
NSPack & 119 & 21MB & PE32 \\
Packman & 115 & 38MB & PE32 \\
PECompact & 124 & 20MB & PE32 \\
PEtite & 124 & 156MB & MSDOS,PE32 \\
RLPack & 115 & 32MB & PE32 \\
Telock & 120 & 21MB & .NET,PE32 \\
Themida & 123 & 361MB & PE32 \\
UPX & 122 & 36MB & PE32 \\
WinUPack & 120 & 20MB & MSDOS,PE32 \\
Yoda's Crypter & 118 & 30MB & .NET,PE32 \\
Yoda's Protector & 115 & 60MB & PE32 \\
Not Packed & 432 & 106MB & PE \\
\caption{Overview of the reference dataset}\label{tab:reference-dataset}
\end{longtable}
\endgroup

\end{document}